\documentclass[pre,aps,twocolumn,showpacs,floats]{revtex4}
\usepackage{graphicx,amsmath,epsf}

\begin{document}
\title{The spectral form factor near the Ehrenfest-time}

\author{Piet W.\ Brouwer and Saar Rahav}

\affiliation{Laboratory of Atomic and Solid State Physics, Cornell
  University, Ithaca 14853, USA}

\date{\today}

\author{Chushun Tian}

\affiliation{Institut f{\"u}r Theoretische Physik, Universit{\"a}t
zu K{\"o}ln, K{\"o}ln, 50937, Germany }

\begin{abstract}
We calculate the Ehrenfest-time dependence
of the leading quantum correction to the
spectral form factor of a ballistic chaotic cavity using
periodic orbit theory. For the case of broken time-reversal
symmetry, our result differs from that previously obtained using
field-theoretic methods [Tian and Larkin, Phys.\ Rev.\ B {\bf 70},
035305 (2004)]. The discrepancy shows that short-time
regularization procedures dramatically affect physics near the
Ehrenfest-time.
\end{abstract}


\maketitle

{}

From a statistical point of view, the spectra of a quantum
particle confined to a cavity with disorder or confined to a
cavity with ballistic and chaotic classical dynamics are
remarkably similar \cite{kn:akkermans1995}. In both cases, the
spectral statistics are independent of system specifics and follow
the predictions of random matrix theory (RMT) on energy scales
comparable to the mean spacing between energy levels $\Delta$ or,
equivalently, on time scales comparable to the Heisenberg time
$t_{\rm H} = 2 \pi \hbar/\Delta$
\cite{kn:efetov1983,kn:bohigas1984}, provided that the latter is
much larger than the time $\tau_{\rm erg}$ needed for ergodic
exploration of the phase space. Calculations of the spectral
statistics have been the theoretical vehicle through which the
profound correspondence between RMT, impurity diagrammatic
perturbation theory, periodic-orbit theory, and the
field-theoretic zero-dimensional $\sigma$ model has been
demonstrated
\cite{kn:efetov1983,kn:smith1998,kn:tian2004b,kn:mueller2004,kn:mueller2005}.

In the last decade, it has been understood that chaotic quantum
systems are characterized by a time scale intermediate between
$\tau_{\rm erg}$ and $\tau_{\rm H}$ that does not have its
counterpart in diffusive systems. This time scale is the
`Ehrenfest time' $\tau_{\rm E}$
\cite{kn:larkin1968,kn:zaslavsky1981,kn:chirikov1981,kn:aleiner1996}.
The Ehrenfest time is the time required for two classical
trajectories initially a quantum distance (wavelength) apart to
diverge and reach a classical separation (system size). It is
expressed in terms of the Lyapunov exponent $\lambda$ of the
corresponding classical dynamics as $\tau_{\rm E} = \lambda^{-1}
\ln (c^2/\hbar)$, where $c^2$ is a classical (action) scale.
The existence of the Ehrenfest time does not affect the universality of
the spectral statistics. However, it is responsible for differences
between otherwise universal properties of chaotic and disordered
quantum systems for
energies $\sim \hbar/\tau_{\rm E}$ , which is well inside the
universal range if $\hbar/c^2 \to 0$
\cite{kn:aleiner1996,kn:aleiner1997,kn:agam2000,kn:vavilov2003,kn:silvestrov2003b,kn:tian2004b}.
In this article we address the spectral statistics at this energy
scale.

Most of the literature on spectral statistics considers the spectral
form factor $K(t)$, which is the Fourier-transform of the two-point
correlation function of the level density $\rho(\varepsilon)$,
\begin{equation}
  K(t) =  \hbar \left\langle \int d\omega
  e^{i \omega t} \rho(\varepsilon + \hbar \omega/2)
  \rho(\varepsilon - \hbar \omega/2) \right\rangle_{\rm c}.
\end{equation}
Here the brackets $\langle \ldots \rangle_{\rm c}$ denote the
connected average obtained by varying the center energy $\varepsilon$
and/or other parameters in the system.
For $\tau_{\rm erg} \ll t < t_{\rm H}$, the RMT predicts that
$K(t)$ is dominated by a perturbative expansion in $t/t_{\rm H}$
\cite{kn:berry1985,kn:akkermans1995,kn:mehta1991}
\begin{eqnarray}
  \label{eq:K}
  K(t) &=& \frac{t}{\beta \pi \hbar} +
  \delta K_{\beta}(t),
\end{eqnarray}
where $\beta=1$ ($2$) in the presence (absence) of time-reversal
 symmetry and
\begin{eqnarray}
  \delta K_1(t) &=& - \frac{t}{\pi \hbar}
  \left( \frac{t}{t_{\rm H}} - \frac{t^2}{t_{\rm H}^2}
  + \ldots \right), \nonumber \\
  \delta K_2(t) &=& 0.
\end{eqnarray}
For times $t \gtrsim t_{\rm H}$ the perturbative expansion in
$t/t_{\rm H}$ breaks down, and $K(t)$ is governed by
non-perturbative contributions \cite{kn:mehta1991}.

The presence of the Ehrenfest time does not affect the leading
contribution to $K(t)$, but it does impact $\delta K$. The leading
$\tau_{\rm E}$ dependence of $\delta K_1$ in the perturbative regime,
which already occurs to order $(t/t_{\rm H})^2$, was first
considered by Aleiner and Larkin \cite{kn:aleiner1997}, using a
field-theoretic approach. Recently, Tian and Larkin used the
field-theoretic
approach to calculate the leading $\tau_{\rm E}$-dependence of
$\delta K_2$ \cite{kn:tian2004b}, which appears to order $(t/t_{\rm
  H})^3$ and causes the perturbative quantum correction $\delta K_2$
no longer to be strictly zero. Their result is
\begin{eqnarray}
  \label{eq:dKK1}
  \delta K_1(t) &=& - \frac{t^2}{\pi \hbar t_{\rm H}}
  \theta(t-2\tau_{\rm E}) + \ldots, \\
  \delta K_2(t) &=& - \frac{t^2}{2 \pi \hbar t_{\rm H}^2}
  \left[ \Theta(t-3 \tau_{\rm E})
  - \Theta(t-4 \tau_{\rm E}) \right]
  + \ldots , ~~~
\end{eqnarray}
where $\theta(x) = 1$ if $x > 0$ and $\theta(x) = 0$ otherwise,
$\Theta(x) = \int^x_0 dx' \theta(x') = x \theta(x)$, and the dots
$\ldots$ refer to terms of higher order in $t/t_{\rm H}$ that were
not considered in the calculation. (Tian and Larkin also
considered the $\tau_{\rm E}$-dependence of non-perturbative
contributions to the spectral form factor, but these will not be
considered here.)

In a parallel development, the full perturbation expansion for $K(t)$
was derived from periodic-orbit
theory
\cite{kn:sieber2001,kn:heusler2004,kn:mueller2004,kn:mueller2005}.
The
connection between $K(t)$, which is a quantum-mechanical object, and
classical periodic orbits follows from Gutzwiller's trace formula,
which expresses $K(t)$ as a double sum over periodic orbits
$\gamma$ and $\gamma'$ \cite{kn:berry1985,kn:gutzwiller1990},
\begin{eqnarray}
  K(t) &=&
  \left\langle
  \sum_{\gamma,\gamma'} A_{\gamma} A_{\gamma'}^*
  e^{i(S_{\gamma} - S_{\gamma'})/\hbar}
  \delta \left(t - \frac{T_{\gamma} + T_{\gamma'}}{2} \right)
  \right\rangle. ~~~~
\end{eqnarray}
Here $A_{\gamma}$, $S_{\gamma}$, and $T_{\gamma}$ are the
stability amplitude, classical action, and period of $\gamma$,
respectively. While the leading contribution to $K(t)$ comes from
the diagonal terms $\gamma = \gamma'$ (up to time-reversal, if
$\beta=1)$ \cite{kn:berry1985}, off-diagonal contributions are
responsible for $\delta K(t)$. Sieber and Richter
\cite{kn:sieber2001} and Heusler {\em et al.}\
\cite{kn:heusler2004,kn:mueller2004,kn:mueller2005} succeeded in
classifying the relevant off-diagonal orbit pairs and calculated
their contribution to $\delta K(t)$ in the limit $\tau_{\rm
E}/t_{\rm H} \to 0$.

Below, we show that periodic-orbit theory can also be used to
calculate the Ehrenfest-time dependence of $\delta K(t)$.
Interestingly, while we confirm the field-theoretic result for the
${\cal O}(t^2)$ term in $\delta K_1(t)$
\cite{kn:tian2004b,kn:altland2005}, our result for the leading
${\cal O}(t^3)$ contribution to $\delta K_2(t)$ differs from that
of Ref.\ \onlinecite{kn:tian2004b},
\begin{eqnarray}
  \label{eq:dKKK2}
  \delta K_2(t) &=&
  \frac{3 t^2}{2 \pi \hbar t_{\rm H}^2}
  \\ && \mbox{} \times
  \left[\Theta(t - 2 \tau_{\rm E}) - 2 \Theta(t - 3 \tau_{\rm E}) +
  \Theta(t - 4 \tau_{\rm E}) \right]. \nonumber
\end{eqnarray}
In particular, we find that the minimum duration of off-diagonal
pairs of orbits that contribute to $\delta K_2$ is $2 \tau_{\rm
E}$, not $3 \tau_{\rm E}$. We also find that there are no
$\tau_{\rm E}$-dependent corrections for $t > 4 \tau_{\rm E}$,
in contrast to Ref.\ \onlinecite{kn:tian2004b}, where the
$\tau_{\rm E}$-dependent corrections to $\delta K_2$ persist up to
$t \sim t_{\rm H}$. (However, our leading-order perturbative calculation
does not answer the question whether such $\tau_{\rm E}$-insensitivity
at long times persist to the higher order terms in $t/t_{\rm H}$.)

Instead of calculating $K(t)$ directly, it is more convenient to
calculate the Laplace transform
\begin{eqnarray}
  K(\alpha) &=&
  \left\langle
  \sum_{\gamma,\gamma'} A_{\gamma} A_{\gamma'}^*
  e^{i(S_{\gamma} - S_{\gamma'})/\hbar
  - \alpha (T_{\gamma} + T_{\gamma'})/2}
  \right\rangle.
\end{eqnarray}
The leading diagonal contribution to $K$
can be calculated using the sum rule of
Hannay and Ozorio de Almeida \cite{kn:hannay1984},
\begin{equation}
  \label{eq:Hannay}
  \left\langle
  \sum_{\gamma} |A_{\gamma}|^2 e^{-\alpha T_{\gamma}}
  \right\rangle
  = \frac{1}{2 \pi \hbar \alpha^2},
\end{equation}
so that
\begin{equation}
  K(\alpha) = \frac{1}{\pi \hbar \alpha^2 \beta} + \delta K(\alpha).
  \label{eq:Kdiagonal}
\end{equation}
The inverse Laplace transform of the first term in
Eq.\ (\ref{eq:Kdiagonal}) reproduces
the leading term in Eq.\ (\ref{eq:K}) above.

\begin{figure}
\epsfxsize=0.99\hsize
\epsffile{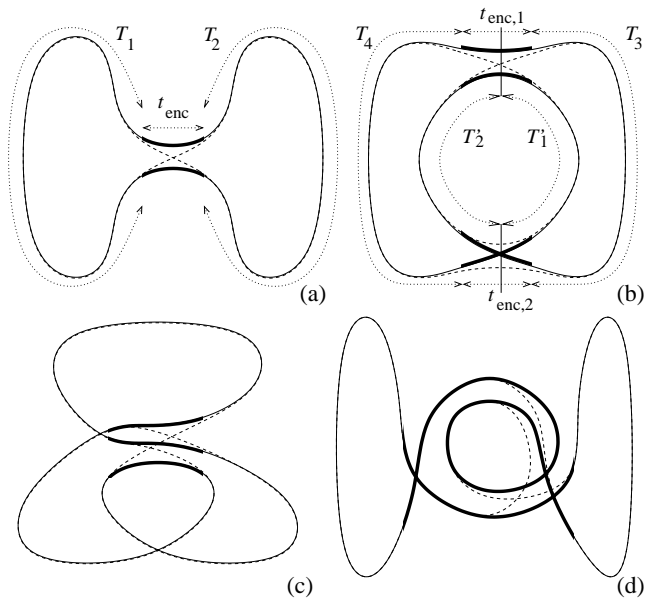}
\caption{\label{fig:1} \label{fig:2}
Schematic drawing of a pair of orbits (shown solid and dashed)
contributing to the leading
  interference correction to the spectral form factor in the
  presence of time-reversal symmetry (a) and in the absence of
  time-reversal symmetry (b,c,d). The true orbits are piecewise
  straight, with specular reflection off the cavity's boundaries.
  The small-angle self-encounters are shown thick. Panels (a) and (b)
  also show the definitions of various durations used in the text.}
\end{figure}

The leading ${\cal O}(t^2)$ quantum correction for $K(\alpha)$
exists in the presence of time-reversal symmetry only. The
relevant pairs of periodic orbits $\gamma$ and $\gamma'$ are shown
in Fig.\ \ref{fig:1}a. The existence of such pairs was pointed out
by Sieber and Richter \cite{kn:sieber2001}; an equivalent
configuration of classical trajectories appears in the
field-theoretic formulation
\cite{kn:aleiner1997,kn:tian2004b,kn:altland2005} and in the
diagrammatic calculation of the form factor for disordered
cavities \cite{kn:smith1998}. The periodic orbit $\gamma$ in Fig.\
\ref{fig:1} has a small-angle self-intersection. There are two
loops of duration $T_1$ and $T_2$ through which $\gamma$ returns
to the self-intersection. The trajectory $\gamma'$ is equal to
$\gamma$ in one of these loops, whereas $\gamma'$ is the
time-reversed of $\gamma$ in the other loop. Following Ref.\
\onlinecite{kn:heusler2004,kn:mueller2004,kn:mueller2005}, we
perform the sum over such periodic orbits with the help of a
Poincar\'e surface of section taken at an arbitrary point during
the self-intersection. The Poincar\'e surface of section is
parameterized using stable and unstable phase space coordinates
$s$ and $u$, normalized such that $ds du$ is the cross-sectional
area element. Denoting the coordinate differences between the two
points where $\gamma$ pierces the Poincar\'e surface of section by
$s$ and $u$, the action difference $S_{\gamma} - S_{\gamma'} = su$
\cite{kn:spehner2003,kn:turek2003}. The duration $t_{\rm enc}$ of
the self-encounter is defined as the time during which the two
stretches of $\gamma$ are within a phase space distance $c$, where
$c$ is a classical scale below which the classical dynamics can be
linearized. The periodic-orbit sum is then expressed in terms of
an integral over $s$, $u$, $T_1$ and $T_2$
\cite{kn:heusler2004,kn:mueller2004,kn:mueller2005},
\begin{eqnarray}
  \label{eq:dK1med}
  \delta K_1(\alpha) &=&
  \int dT_1 dT_2 \int_{-c}^{c} ds du
  \frac{(T_1+T_2+2 t_{\rm enc})^2}{(2 \pi \hbar)^2 t_{\rm H} t_{\rm enc}}
  \nonumber \\ && \mbox{} \times
  \cos(s u/\hbar)
  e^{-\alpha(T_1+T_2+2 t_{\rm enc})}
  \nonumber \\ &=&
  \frac{\partial^2}{\partial
  \alpha^2}
  \frac{1}{\alpha^2}
  \int_{-c}^{c} ds du \frac{\cos(s u/\hbar) e^{-2 \alpha t_{\rm
  enc}}}{(2 \pi \hbar)^2 t_{\rm H} t_{\rm enc}},
\end{eqnarray}
where
\begin{equation}
  t_{\rm enc} = (1/\lambda) \ln(c^2/|su|),
\end{equation}
$\lambda$ being the Lyapunov exponent of the classical dynamics in
the cavity. The factor $t_{\rm enc}$ in the denominator cancels an
unwanted contribution from the freedom to choose the Poincar\'e
surface of section anywhere inside the encounter region. Whereas
Refs.\ \onlinecite{kn:heusler2004,kn:mueller2004,kn:mueller2005}
calculate the remaining integral in the limit $\alpha \tau_{\rm E}
\to 0$, we need to consider the effect of a finite Ehrenfest time.
Such integrals have been considered in the context of quantum
transport and we will be able to obtain the remaining integral in
Eq.\ (\ref{eq:dK1med}) as well as all other necessary integrals
from the literature. The integral needed here has been calculated
in Ref.\ \onlinecite{kn:brouwer2006} and gives
\begin{equation}
  \delta K_1(\alpha) =
  -\frac{1}{\pi \hbar t_{\rm H}}
  \frac{\partial^2}{\partial
  \alpha^2} \frac{e^{-2 \alpha \tau_{\rm E}}}{\alpha}.
  \label{eq:dk1}
\end{equation}
where
\begin{equation}
  \tau_{\rm E} = \frac{1}{\lambda} \ln \frac{c^2}{\hbar}.
\end{equation}
This result is in agreement with the $\tau_{\rm E}$ dependence of
$\delta K_1$ calculated by Tian and Larkin
\cite{kn:tian2004b}. Its inverse Laplace transform is Eq.\
(\ref{eq:dKK1}) above.

We now consider the Ehrenfest-time dependent correction $\delta
K_2(t)$ to the spectral form factor in the absence of
time-reversal symmetry. There are three classes of periodic orbits
that give a contribution to $\delta K_2(t)$ to order $t^3$. These
are shown in Fig.\ \ref{fig:2}b--d. Now, the classical orbit
$\gamma$ has two small-angle self encounters. Its partner orbit
$\gamma'$ follows $\gamma$ between the encounters, but connects
the ends of the encounters in a different way. Figure \ref{fig:2}b
shows two separate encounters. Figure \ref{fig:2}c shows a
``three-encounter'', which arises when the two small-angle
encounters of Fig.\ \ref{fig:2}b are merged along one of the
connecting stretches (while keeping the duration of the other
stretches finite). The orbit $\gamma$ of Fig.\ \ref{fig:2}c goes
through three loops between returns to the encounter region.
Finally, Fig.\ \ref{fig:2}d shows a periodic orbit for which the
self-encounter fully extends along one of these loops
\cite{kn:fnbrouwer2006}.
Note that the encounter region can not
simultaneously extend along two or more of the loops, because then
these loops, and, hence, the trajectories $\gamma$ and $\gamma'$
would be equal.

The configurations of Fig.\ \ref{fig:2}b and c were also
considered by Heusler {\em et al.}
\cite{kn:heusler2004,kn:mueller2004,kn:mueller2005}. These two
contributions cancel in the limit $\alpha \tau_{\rm E} \to 0$, so
that one finds $\delta K_2 = 0$ in that limit. The contribution of
the trajectories shown Fig.\ \ref{fig:2}d vanishes in the limit
$\alpha \tau_{\rm E} \to 0$, which is why it was not considered in
Refs.\ \onlinecite{kn:heusler2004,kn:mueller2004,kn:mueller2005}.
However, as we will show below, it is needed to be taken into
account when calculating $\tau_{\rm
  E}$-dependent corrections. The field-theoretic calculation of
Ref.\ \onlinecite{kn:tian2004b} also has two contributions to
$\delta K_2$ only, one with two two-encounters and the other with
a single three-encounter.

The contribution $\delta K_{2b}$ of the two separate encounters for the
periodic-orbit pair in Fig.\ \ref{fig:2}b factorizes. Taking a
Poincar\'e surface of section at each of the encounters and
proceeding as in the calculation of $\delta K_1$, one finds
\begin{eqnarray}
  \label{eq:dK2a}
  \delta K_{2b}(\alpha) &=& \frac{1}{8 \pi \hbar}
  \frac{\partial^2}{\partial \alpha^2}
  \frac{1}{\alpha^4}
  \left[\int_{-c}^{c} ds du \frac{\cos(s u/\hbar) e^{-2 \alpha t_{\rm
  enc}}}{2 \pi \hbar t_{\rm H} t_{\rm enc}} \right]^2
  \nonumber \\ &=&
  \frac{1}{2 \pi \hbar t_{\rm H}^2}
  \frac{\partial^2}{\partial \alpha^2} \frac{e^{-4 \alpha \tau_{\rm
  E}}}{\alpha^2},
\end{eqnarray}
where we included a combinatorial factor $1/4$ to account for
permutations of the $2 \times 2$ passages of $\gamma$ through the
two encounters
\cite{kn:heusler2004,kn:mueller2004,kn:mueller2005}.
This result is in agreement with that obtained by calculating the
two two-encounter diagram in the field-theoretic approach
\cite{kn:foottian2004b}.

For the calculation of the contribution of the trajectory of
Fig.\ \ref{fig:2}c one needs only one Poincar\'e surface of section,
taken at a point where all three stretches of $\gamma$ are within a
phase space distance $c$. Labeling the phase space coordinates of the
three piercings of $\gamma$ through the surface of section as
$(s_i,u_i)$, $i=1,2,3$, the action difference is
\cite{kn:heusler2004,kn:mueller2005}
\begin{equation}
  S_{\gamma} - S_{\gamma'} = s u + s' u',
\end{equation}
where $s = s_1-s_3$, $s' = s_1-s_5$, $u = u_1-u_3$,
and $u' = u_1-u_5$. Integrating over the durations of the three stretches
of $\gamma$ that connect the three-encounter to itself, we find
\begin{eqnarray}
  \label{eq:dK2}
  \delta K_{2c}(\alpha) &=&
  \frac{1}{3} \frac{\partial^2}{\partial \alpha^2}
  \frac{1}{\alpha^2}
  \int ds ds' du du'
  \nonumber \\ && \mbox{} \times
  \frac{\cos[(su + s'u')/\hbar]\, e^{-\alpha (2t_{\rm enc} +
  t_{\rm enc}'})}{(2 \pi \hbar)^3 t_{\rm H}^2 t_{\rm enc}'},
\end{eqnarray}
where $t_{{\rm enc}}$ is the duration of the encounter,
\begin{eqnarray*}
  t_{{\rm enc}} &=&
    \frac{1}{\lambda} \ln \frac{c^2}{\min(|s|,|s'|,|s-s'|)
    \min(|u|,|u'|,|u+u'|)}
\end{eqnarray*}
and $t_{{\rm enc}}'$ is the time that all three
trajectories involved in the encounter are within a phase space
distance $c$,
\begin{eqnarray*}
  t_{{\rm enc}}' &=&
    \frac{1}{\lambda} \ln \frac{c^2}{\max(|s|,|s'|,|s-s'|)
    \max(|u|,|u'|,|u+u'|)}.
\end{eqnarray*}
The prefactor $1/3$ in Eq.\
(\ref{eq:dK2}) accounts for permutations of the three passages of
$\gamma$ through the three encounter.
The integration domain in Eq.\ (\ref{eq:dK2}) is
$\max(|u|,|u'|,|u+u'|),\max(|s|,|s'|,|s-s'|) < c$. The exponential
factor contains the total time $2 t_{\rm enc} + t_{\rm enc}'$ that the
orbit $\gamma$ spends in the encounter region. Taking the remaining
integral over $s$, $s'$, $u$, and $u'$ from Ref.\
\onlinecite{kn:brouwer2006c}, we find
\begin{eqnarray}
  \delta K_{2c}(\alpha) &=&
  \frac{1}{2 \pi \hbar t_{\rm H}^2}
  \frac{\partial^2}{\partial \alpha^2}
  \frac{1}{\alpha^2}
  \left[3 e^{-3 \alpha \tau_{\rm E}} - 4 e^{-4 \alpha \tau_{\rm E}} \right].
\end{eqnarray}
Although this result agrees with the field-theoretic calculation
in the limit $\alpha\tau_{\rm E}\rightarrow 0$
\cite{kn:smith1998,kn:tian2004b}, the appearance of a term
proportional to $\exp(-4 \alpha \tau_{\rm E})$-oscillation differs
from the field-theoretic approach \cite{kn:tian2004b}, in which
only one term proportional to $\exp(-3 \alpha \tau_{\rm E})$
appears with prefactor $-1$ (=3-4).

Finally, we have to calculate the contribution from trajectories with
a three-encounter where the encounter region fully wraps around one of
the loops. In order to make optimal use of the literature on the
Ehrenfest-time dependence of quantum transport, we calculate this
contribution in an indirect way: We again consider the case of
two two-encounters, but now
allow the encounters to approach each other and overlap along two
pre-assigned stretches of $\gamma$. This situation shown in
Fig.\ \ref{fig:2}b (again), and we allow the encounters to approach each other
along the central loop in the figure. We take a Poincar\'e surface of
section at each encounter, and measure the durations of between the
two surfaces of section along the central loop by $T_1'$ and $T_2'$.
Along the remaining stretches (the outer loop in Fig.\ \ref{fig:2}b),
we still require non-overlapping encounters in order to enforce
$\gamma \neq \gamma'$. Hence, we parameterize their
duration using times $T_3$ and $T_4$ measured between the ends of the
encounters. Finally, $t_{{\rm enc},1}$ and $t_{{\rm enc},2}$ denote
time that the inner and outer loops are within a phase space distance
$c$, and $t_s$ and $t_u$ are the durations of eventual stretches that
the two segments of the outer loop are within a phase space distance
$c$ from each other but not from the inner loop \cite{kn:brouwer2006}.
(The times $t_s$ and
$t_u$ are zero except in the case of overlapping encounters.) With this
parameterization, the total duration of $\gamma$ is
\begin{equation}
  T_{\gamma} = T_1' + T_2' + T_3 + T_4 + t_{{\rm enc},1} + t_{{\rm
  enc},2} + 2 t_s + 2 t_u.
\end{equation}

We now consider the integral
\begin{eqnarray}
  I &=&
  \int dT_1' dT_2' dT_3 dT_4
  \int_{-c}^{c} ds_1 du_1 ds_2 du_2
  \nonumber \\ && \mbox{} \times
  \frac{T_{\gamma}^2
  \cos[(u_1 s_1 + u_2 s_2)/\hbar]   e^{-\alpha T_{\gamma}}
}{(2 \pi \hbar)^3 t_{\rm H}^2
  t_{{\rm enc},1} t_{{\rm enc},2}},
  \label{eq:I}
\end{eqnarray}
where $s_i$ and $u_i$ are phase space coordinates at the two
Poincar\'e surfaces of section, $i=1,2$. This integral contains
both the case that the two encounters are separate and the case that
the two encounters overlap. If the two encounters are separate,
which requires $T_1' > \lambda^{-1} \ln (|u_1 s_2|/c^2)$ and $T_2' >
\lambda^{-1} \ln(|u_2 s_1|/c^2)$, the two stretches in the outer loop
are never close to each other, hence $t_s = t_u = 0$. One then
recovers the expression for $\delta K_{2b}$, multiplied by four because
of the combinatorial factor $1/4$ which is present in Eq.\
(\ref{eq:dK2a}) but not in Eq.\ (\ref{eq:I}). If the two
encounters overlap at one end but not at the other end, one recovers
the scenario for $\delta K_{2c}$, multiplied by three because of the
combinatorial factor $1/3$ which is present in Eq.\ (\ref{eq:dK2}) but
not in Eq.\ (\ref{eq:I}). (Note that in this case
the times $t_s$ and $t_u$ need not be zero.) If the
encounters overlap at two ends, they span the reference loop. This
scenario is neither contained in $\delta K_{2b}$ nor in $\delta
K_{2c}$. Since there is no combinatorial factor in this case, this is
precisely the contribution $\delta K_{2d}$ corresponding to the
trajectories of the type shown in Fig.\ \ref{fig:2}d. Hence
\begin{equation}
  \delta K_{2d}(\alpha) = I - 4 \delta K_{2b}(\alpha) - 3 \delta
  K_{2c}(\alpha).
\end{equation}
{}From Sec.\ IV of Ref.\ \onlinecite{kn:brouwer2006},
where an integral similar to $I$ was calculated, we find
\begin{equation}
  I = \frac{1}{2 \pi \hbar t_{\rm H}^2}
  \frac{\partial^2}{\partial \alpha^2} \frac{1}{\alpha^2}
  \left[3 e^{-2\alpha \tau_{\rm E}} - 2 e^{-4 \alpha \tau_{\rm E}} \right].
  \label{eq:Iresult}
\end{equation}
Combining everything, we arrive at
\begin{eqnarray}
  \delta K_2(\alpha) &=&
  \delta K_{2b}(\alpha) + \delta K_{2c}(\alpha) + \delta
  K_{2d}(\alpha)
  \nonumber \\ &=&
  \frac{3}{2 \pi \hbar t_{\rm H}^2}
  \frac{\partial^2}{\partial \alpha^2}
  \frac{e^{-2 \alpha \tau_{\rm E}}}{\alpha^2}
  \left[1-e^{-\alpha \tau_{\rm E}}\right]^2.
  \label{eq:dKfinal}
\end{eqnarray}
The inverse Laplace transform of this result is Eq.\ (\ref{eq:dKKK2})
above.

The main difference between our result and that previously
obtained by the field-theoretic approach is that, in contrast to
Ref.\ \onlinecite{kn:tian2004b}, in Eq.\ (\ref{eq:dKKK2})
universal quantum corrections already appear after a time $2
\tau_{\rm E}$ \cite{kn:fntian2006}.
This shortest-duration contribution to $\delta K_2$ stems from
periodic orbit pairs which (for a part of their duration) wind
around another, shorter, periodic orbit. Such orbits
explained \cite{kn:brouwer2006} the numerically observed
$\tau_{\rm E}$-independence of conductance fluctuations in chaotic
cavities \cite{kn:tworzydlo2004,kn:jacquod2004}. Another
difference between our result and that previously obtained by the
field-theoretic approach is the appearance of a discontinuity at
$4 \tau_{\rm E}$ for the contribution $\delta K_{2c}$ of a single
three encounter. A feature at $4 \tau_{\rm E}$, which is absent in
the field-theoretic calculation, is essential for the validity of
the `effective random matrix theory' \cite{kn:silvestrov2003b} of
the Ehrenfest-time dependence of the spectral gap in a chaotic
cavity coupled to a superconductor \cite{kn:brouwer2006c}.

The calculation of the $\tau_{\rm E}$-dependence of the
perturbative correction $\delta K_2$ is, to the best of our
knowledge, the first example in which the periodic orbit and the
field-theoretical approaches lead to different results in the
universal range. It is also the first example in which the two
approaches are compared for a quantity that involves more than a
single two-encounter. We believe this discrepancy can be traced to
a difference in the way both approaches handle the encounter
regions, in particular to differences implicitly or explicitly
imposed by the short-time regularization procedures.

In the periodic-orbit approach, a generic phase-space density
function is described in reference to a classical trajectory. It
expands and contracts along the unstable and stable directions in
phase space, but with a uniform density function at each point
during the encounter. The total duration of the encounter region
is $\sim \tau_{\rm E}$, beyond which the density functions are
projected onto the low-lying sectors of the Perron-Frobenius modes
and evolve irreversibly \cite{kn:roberts2000}, despite the
deterministic nature of the classical dynamics. On the other hand,
in the field-theoretic approach of Ref.\
\onlinecite{kn:tian2004b}, each `orbit' represents the evolution
from a singular density field concentrated at a point in phase
space. Under the action of elliptical regulators, such singular
fields first become locally smooth, spreading over the scale
determined by the regulator strength. This scale is adjusted to be
$\sim \hbar$ in order to mimic the evolution of quantum wave
packets with minimal variance
\cite{kn:tian2004b,kn:aleiner1996,kn:vavilov2003}. At the time
$\tau_{\rm E}/2$ away from the origin (both forward and backward
in time), the (now) locally smooth fields are projected onto the
low-lying sectors of the Perron-Frobenius modes
\cite{kn:tian2005}. The two approaches give the same result for a
single two-encounter: in both cases the encounter region is a
generic classical trajectory of duration $\tau_{\rm E}$. However,
whether encounters are built from a reference trajectory or a
reference `point' matters for encounters of more than two
trajectories. In the periodic-orbit approach, a reference orbit
can be a periodic orbit with an arbitrarily short period. These
reference orbits then give the universal contribution to $\delta
K$ at the time $2 \tau_{\rm E}$ that was calculated above. The
reference orbit can also be a trajectory of length $\lesssim 2
\tau_{\rm E}$, which is involved in two almost non-overlapping two
encounters. This case gives the feature at time $4 \tau_{\rm E}$
for the three-encounter contribution to the form factor. In the
field-theoretic approach of Ref.\ \onlinecite{kn:tian2004b}, the
evolution of the phase space density field is always altered
irreversibly a time $\tau_{\rm E}/2$ away from the reference
point, thus, in the case of a three-encounter, leaving no room for
a quantum correction at time $2 \tau_{\rm E}$, nor for a feature
at $4 \tau_{\rm E}$.

Summarizing, we used periodic orbit theory to calculate the
leading Ehrenfest time dependent corrections to the spectral form
factor $K(t)$. We find that, in the absence of time-reversal
symmetry, the corrections to the leading RMT prediction do not
strictly vanish for $2 \tau_{\rm E} < t < 4 \tau_{\rm E}$ if the
Ehrenfest time $\tau_{\rm E}$ is finite. The fact that
finite-$\tau_{\rm E}$ corrections give a nonzero correction to
$K(t)$ is in agreement with a previous field-theoretic calculation
\cite{kn:tian2004b}, although our detailed expression for $K(t)$
is not. We attribute the difference to the different short-time
regularization used in Ref.\ \onlinecite{kn:tian2004b}.

We are grateful to A.~Altland and, especially, J.~M{\"u}ller for
important discussions. This work was supported by the NSF under
grant no.\ DMR 0334499, the Packard Foundation and Transregio SFB
12 of the Deutsche Forschungsgemeinschaft.


\vspace{0cm}

\begin{thebibliography}{26}
\expandafter\ifx\csname natexlab\endcsname\relax\def\natexlab#1{#1}\fi
\expandafter\ifx\csname bibnamefont\endcsname\relax
  \def\bibnamefont#1{#1}\fi
\expandafter\ifx\csname bibfnamefont\endcsname\relax
  \def\bibfnamefont#1{#1}\fi
\expandafter\ifx\csname citenamefont\endcsname\relax
  \def\citenamefont#1{#1}\fi
\expandafter\ifx\csname url\endcsname\relax
  \def\url#1{\texttt{#1}}\fi
\expandafter\ifx\csname urlprefix\endcsname\relax\def\urlprefix{URL }\fi
\providecommand{\bibinfo}[2]{#2}
\providecommand{\eprint}[2][]{\url{#2}}

\bibitem[{\citenamefont{Akkermans et~al.}(1995)\citenamefont{Akkermans,
  Montambaux, Pichard, and Zinn-Justin}}]{kn:akkermans1995}
\bibinfo{editor}{\bibfnamefont{E.}~\bibnamefont{Akkermans}},
  \bibinfo{editor}{\bibfnamefont{G.}~\bibnamefont{Montambaux}},
  \bibinfo{editor}{\bibfnamefont{J.-L.} \bibnamefont{Pichard}},
  \bibnamefont{and}
  \bibinfo{editor}{\bibfnamefont{J.}~\bibnamefont{Zinn-Justin}}, eds.,
  \emph{\bibinfo{title}{Mesoscopic Quantum Physics}}
  (\bibinfo{publisher}{North-Holland}, \bibinfo{year}{1995}).

\bibitem[{\citenamefont{Efetov}(1983)}]{kn:efetov1983}
\bibinfo{author}{\bibfnamefont{K.~B.} \bibnamefont{Efetov}},
  \bibinfo{journal}{Adv. Phys.} \textbf{\bibinfo{volume}{32}},
  \bibinfo{pages}{53} (\bibinfo{year}{1983}).

\bibitem[{\citenamefont{Bohigas et~al.}(1984)\citenamefont{Bohigas, Giannoni,
  and Schmit}}]{kn:bohigas1984}
\bibinfo{author}{\bibfnamefont{O.}~\bibnamefont{Bohigas}},
  \bibinfo{author}{\bibfnamefont{M.-J.} \bibnamefont{Giannoni}},
  \bibnamefont{and} \bibinfo{author}{\bibfnamefont{C.}~\bibnamefont{Schmit}},
  \bibinfo{journal}{Phys. Rev. Lett.} \textbf{\bibinfo{volume}{52}},
  \bibinfo{pages}{1} (\bibinfo{year}{1984}).

\bibitem[{\citenamefont{Smith et~al.}(1998)\citenamefont{Smith, Lerner, and
  Altshuler}}]{kn:smith1998}
\bibinfo{author}{\bibfnamefont{R.~A.} \bibnamefont{Smith}},
  \bibinfo{author}{\bibfnamefont{I.~V.} \bibnamefont{Lerner}},
  \bibnamefont{and} \bibinfo{author}{\bibfnamefont{B.~L.}
  \bibnamefont{Altshuler}}, \bibinfo{journal}{Phys. Rev. B}
  \textbf{\bibinfo{volume}{58}}, \bibinfo{pages}{10343} (\bibinfo{year}{1998}).

\bibitem[{\citenamefont{Tian and Larkin}(2004)}]{kn:tian2004b}
\bibinfo{author}{\bibfnamefont{C.}~\bibnamefont{Tian}} \bibnamefont{and}
  \bibinfo{author}{\bibfnamefont{A.~I.} \bibnamefont{Larkin}},
  \bibinfo{journal}{Phys. Rev. B} \textbf{\bibinfo{volume}{70}},
  \bibinfo{eid}{035305} (\bibinfo{year}{2004}).


\bibitem[{\citenamefont{M\"uller et~al.}(2004)\citenamefont{M\"uller, Heusler,
  Braun, Haake, and Altland}}]{kn:mueller2004}
\bibinfo{author}{\bibfnamefont{S.}~\bibnamefont{M\"uller}},
  \bibinfo{author}{\bibfnamefont{S.}~\bibnamefont{Heusler}},
  \bibinfo{author}{\bibfnamefont{P.}~\bibnamefont{Braun}},
  \bibinfo{author}{\bibfnamefont{F.}~\bibnamefont{Haake}}, \bibnamefont{and}
  \bibinfo{author}{\bibfnamefont{A.}~\bibnamefont{Altland}},
  \bibinfo{journal}{Phys. Rev. Lett.} \textbf{\bibinfo{volume}{93}},
  \bibinfo{eid}{014103} (\bibinfo{year}{2004}).

\bibitem[{\citenamefont{M\"uller et~al.}(2005)\citenamefont{M\"uller, Heusler,
  Braun, Haake, and Altland}}]{kn:mueller2005}
\bibinfo{author}{\bibfnamefont{S.}~\bibnamefont{M\"uller}},
  \bibinfo{author}{\bibfnamefont{S.}~\bibnamefont{Heusler}},
  \bibinfo{author}{\bibfnamefont{P.}~\bibnamefont{Braun}},
  \bibinfo{author}{\bibfnamefont{F.}~\bibnamefont{Haake}}, \bibnamefont{and}
  \bibinfo{author}{\bibfnamefont{A.}~\bibnamefont{Altland}},
  \bibinfo{journal}{Phys.\ Rev.\ E} \textbf{\bibinfo{volume}{72}},
  \bibinfo{eid}{046207} (\bibinfo{year}{2005}).

\bibitem[{\citenamefont{Larkin and Ovchinnikov}(1968)}]{kn:larkin1968}
\bibinfo{author}{\bibfnamefont{A.~I.} \bibnamefont{Larkin}} \bibnamefont{and}
  \bibinfo{author}{\bibfnamefont{Y.~N.} \bibnamefont{Ovchinnikov}},
  \bibinfo{journal}{Zh. Eksp. Teor. Fiz.} \textbf{\bibinfo{volume}{55}},
  \bibinfo{pages}{2262} (\bibinfo{year}{1968}) \bibinfo{note}{[Sov.\ Phys.\
  JETP {\bf 28}, 1200 (1969)]}.

\bibitem[{\citenamefont{Zaslavsky}(1981)}]{kn:zaslavsky1981}
\bibinfo{author}{\bibfnamefont{G.~M.} \bibnamefont{Zaslavsky}},
  \bibinfo{journal}{Phys. Rep.} \textbf{\bibinfo{volume}{80}},
  \bibinfo{pages}{157} (\bibinfo{year}{1981}).

\bibitem[{\citenamefont{Chirikov et al}(1981)}]{kn:chirikov1981}
  \bibinfo{author}{\bibfnamefont{B.~V.} \bibnamefont{Chirikov}},
  \bibinfo{author}{\bibfnamefont{F.~M.} \bibnamefont{Izrailev}},
 \bibnamefont{and}
  \bibinfo{author}{\bibfnamefont{D.~L.} \bibnamefont{Shepelyansky}},
  \bibinfo{journal}{Sov. Sci. Rev., Sect. C, Math. Phys. Rev. } \textbf{\bibinfo{volume}{2}},
  \bibinfo{pages}{209} (\bibinfo{year}{1981}).

\bibitem[{\citenamefont{Aleiner and Larkin}(1996)}]{kn:aleiner1996}
\bibinfo{author}{\bibfnamefont{I.~L.} \bibnamefont{Aleiner}} \bibnamefont{and}
  \bibinfo{author}{\bibfnamefont{A.~I.} \bibnamefont{Larkin}},
  \bibinfo{journal}{Phys. Rev. B} \textbf{\bibinfo{volume}{54}},
  \bibinfo{pages}{14423} (\bibinfo{year}{1996}).

\bibitem[{\citenamefont{Aleiner and Larkin}(1997)}]{kn:aleiner1997}
\bibinfo{author}{\bibfnamefont{I.~L.} \bibnamefont{Aleiner}} \bibnamefont{and}
  \bibinfo{author}{\bibfnamefont{A.~I.} \bibnamefont{Larkin}},
  \bibinfo{journal}{Phys. Rev. E} \textbf{\bibinfo{volume}{55}},
  \bibinfo{pages}{R1243} (\bibinfo{year}{1997}). Note that a mistake in this work was corrected in
  Ref.\ \onlinecite{kn:tian2004b}.

\bibitem[{\citenamefont{Agam et~al.}(2000)\citenamefont{Agam, Aleiner, and
  Larkin}}]{kn:agam2000}
\bibinfo{author}{\bibfnamefont{O.}~\bibnamefont{Agam}},
  \bibinfo{author}{\bibfnamefont{I.}~\bibnamefont{Aleiner}}, \bibnamefont{and}
  \bibinfo{author}{\bibfnamefont{A.}~\bibnamefont{Larkin}},
  \bibinfo{journal}{Phys. Rev. Lett.} \textbf{\bibinfo{volume}{85}},
  \bibinfo{pages}{3153} (\bibinfo{year}{2000}).

\bibitem[{\citenamefont{Vavilov and Larkin}(2003)}]{kn:vavilov2003}
\bibinfo{author}{\bibfnamefont{M.~G.} \bibnamefont{Vavilov}} \bibnamefont{and}
  \bibinfo{author}{\bibfnamefont{A.~I.} \bibnamefont{Larkin}},
  \bibinfo{journal}{Phys. Rev. B} \textbf{\bibinfo{volume}{67}},
  \bibinfo{eid}{115335} (\bibinfo{year}{2003}).

\bibitem[{\citenamefont{Silvestrov et~al.}(2003)\citenamefont{Silvestrov,
  Goorden, and Beenakker}}]{kn:silvestrov2003b}
\bibinfo{author}{\bibfnamefont{P.~G.} \bibnamefont{Silvestrov}},
  \bibinfo{author}{\bibfnamefont{M.~C.} \bibnamefont{Goorden}},
  \bibnamefont{and} \bibinfo{author}{\bibfnamefont{C.~W.~J.}
  \bibnamefont{Beenakker}}, \bibinfo{journal}{Phys. Rev. Lett.}
  \textbf{\bibinfo{volume}{90}}, \bibinfo{eid}{116801} (\bibinfo{year}{2003}).

\bibitem[{\citenamefont{Mehta}(1991)}]{kn:mehta1991}
\bibinfo{author}{\bibfnamefont{M.~L.} \bibnamefont{Mehta}},
  \emph{\bibinfo{title}{Random Matrices}} (\bibinfo{publisher}{Academic Press},
  \bibinfo{year}{1991}), \bibinfo{edition}{2nd} ed.

\bibitem[{\citenamefont{Berry}(1985)}]{kn:berry1985}
\bibinfo{author}{\bibfnamefont{M.~V.} \bibnamefont{Berry}},
  \bibinfo{journal}{Proc. R. Soc. London A} \textbf{\bibinfo{volume}{400}},
  \bibinfo{pages}{229} (\bibinfo{year}{1985}).

\bibitem[{\citenamefont{Sieber and Richter}(2001)}]{kn:sieber2001}
\bibinfo{author}{\bibfnamefont{M.}~\bibnamefont{Sieber}} \bibnamefont{and}
  \bibinfo{author}{\bibfnamefont{K.}~\bibnamefont{Richter}},
  \bibinfo{journal}{Phys. Scripta} \textbf{\bibinfo{volume}{T90}},
  \bibinfo{pages}{128} (\bibinfo{year}{2001}).

\bibitem[{\citenamefont{Heusler et~al.}(2004)\citenamefont{Heusler, M\"uller,
  Braun, and Haake}}]{kn:heusler2004}
\bibinfo{author}{\bibfnamefont{S.}~\bibnamefont{Heusler}},
  \bibinfo{author}{\bibfnamefont{S.}~\bibnamefont{M\"uller}},
  \bibinfo{author}{\bibfnamefont{P.}~\bibnamefont{Braun}}, \bibnamefont{and}
  \bibinfo{author}{\bibfnamefont{F.}~\bibnamefont{Haake}},
  \bibinfo{journal}{J.\ Phys. A} \textbf{\bibinfo{volume}{37}},
  \bibinfo{pages}{L31} (\bibinfo{year}{2004}).

\bibitem[{\citenamefont{Gutzwiller}(1990)}]{kn:gutzwiller1990}
\bibinfo{author}{\bibfnamefont{M.}~\bibnamefont{Gutzwiller}},
  \emph{\bibinfo{title}{Chaos in Classical and Quantum Mechanics}}
  (\bibinfo{publisher}{Springer, New York}, \bibinfo{year}{1990}).

\bibitem[{\citenamefont{Mueller and Altland}(2005)\citenamefont{mueller}}]{kn:altland2005}
\bibinfo{author}{\bibfnamefont{J.} \bibnamefont{M{\"u}ller}}
  \bibnamefont{and} \bibinfo{author}{\bibfnamefont{A.}
  \bibnamefont{Altland}}, \bibinfo{journal}{J. Phys. A}
  \textbf{\bibinfo{volume}{38}}, \bibinfo{eid}{3097} (\bibinfo{year}{2005}).

\bibitem[{\citenamefont{Hannay and de~Almeida}(1984)}]{kn:hannay1984}
\bibinfo{author}{\bibfnamefont{J.~H.} \bibnamefont{Hannay}} \bibnamefont{and}
  \bibinfo{author}{\bibfnamefont{A.~M.~O.} \bibnamefont{de~Almeida}},
  \bibinfo{journal}{J. Phys. A} \textbf{\bibinfo{volume}{17}},
  \bibinfo{pages}{3429} (\bibinfo{year}{1984}).

\bibitem[{\citenamefont{Spehner}(2003)}]{kn:spehner2003}
\bibinfo{author}{\bibfnamefont{D.}~\bibnamefont{Spehner}}, \bibinfo{journal}{J.
  Phys. A} \textbf{\bibinfo{volume}{36}}, \bibinfo{pages}{7269}
  (\bibinfo{year}{2003}).

\bibitem[{\citenamefont{Turek and Richter}(2003)}]{kn:turek2003}
\bibinfo{author}{\bibfnamefont{M.}~\bibnamefont{Turek}} \bibnamefont{and}
  \bibinfo{author}{\bibfnamefont{K.}~\bibnamefont{Richter}},
  \bibinfo{journal}{J. Phys. A} \textbf{\bibinfo{volume}{36}},
  \bibinfo{pages}{L455} (\bibinfo{year}{2003}).

\bibitem[{\citenamefont{Brouwer and Rahav}(2005)}]{kn:brouwer2006}
\bibinfo{author}{\bibfnamefont{P.~W.} \bibnamefont{Brouwer}} \bibnamefont{and}
  \bibinfo{author}{\bibfnamefont{S.}~\bibnamefont{Rahav}},
  \bibinfo{journal}{cond-mat/0512095}  (\bibinfo{year}{2005}).

\bibitem[{\citenamefont{footnote Brouwer}(2006)}]{kn:fnbrouwer2006}
In contrast to Refs.\
\onlinecite{kn:heusler2004,kn:mueller2004,kn:mueller2005}, we here
define the encounter region as the total stretch along $\gamma$ at
which at least two stretches of $\gamma$ are within a phase space
distance $c$.

\bibitem{kn:foottian2004b}
The agreement of $\delta K_{1,2b}(\alpha)$ with the corresponding
field-theoretic contribution to $\delta K_2$ 
justifies the crucial technical approximation not to retain
the Liouville operators appearing at vertices throughout the
field-theoretic calculation of Ref.\ \onlinecite{kn:tian2004b},
see Sec. II of Ref.\ \onlinecite{kn:tian2004b}.

\bibitem[{\citenamefont{Brouwer and Rahav}(2006)}]{kn:brouwer2006c}
\bibinfo{author}{\bibfnamefont{P.~W.} \bibnamefont{Brouwer}} \bibnamefont{and}
  \bibinfo{author}{\bibfnamefont{S.}~\bibnamefont{Rahav}},
  \bibinfo{journal}{cond-mat/0606384}  (\bibinfo{year}{2006}).

\bibitem[{\citenamefont{footnote Tian}(2006)}]{kn:fntian2006}
We should point out that the field-theoretic approach of Ref.
\onlinecite{kn:tian2004b} does give a smaller and nonuniversal
quantum correction to $\delta K_2 (t)$ at $t = 2\tau_{\rm E}$.


\bibitem[{\citenamefont{Tworzydlo et~al.}(2004)\citenamefont{Tworzydlo, Tajic,
  and Beenakker}}]{kn:tworzydlo2004}
\bibinfo{author}{\bibfnamefont{J.}~\bibnamefont{Tworzydlo}},
  \bibinfo{author}{\bibfnamefont{A.}~\bibnamefont{Tajic}}, \bibnamefont{and}
  \bibinfo{author}{\bibfnamefont{C.~W.~J.} \bibnamefont{Beenakker}},
  \bibinfo{journal}{Phys. Rev. B} \textbf{\bibinfo{volume}{69}},
  \bibinfo{eid}{165318} (\bibinfo{year}{2004}).

\bibitem[{\citenamefont{Jacquod and Sukhorukov}(2004)}]{kn:jacquod2004}
\bibinfo{author}{\bibfnamefont{P.}~\bibnamefont{Jacquod}} \bibnamefont{and}
  \bibinfo{author}{\bibfnamefont{E.~V.} \bibnamefont{Sukhorukov}},
  \bibinfo{journal}{Phys. Rev. Lett.} \textbf{\bibinfo{volume}{92}},
  \bibinfo{eid}{116801} (\bibinfo{year}{2004}).

\bibitem[{\citenamefont{Roberts and Muzykantskii}(2000)}]{kn:roberts2000}
\bibinfo{author}{\bibfnamefont{S.} \bibnamefont{Roberts}} \bibnamefont{and}
  \bibinfo{author}{\bibfnamefont{B.}~\bibnamefont{Muzykantskii}},
  \bibinfo{journal}{J. Phys. A} \textbf{\bibinfo{volume}{33}},
  \bibinfo{pages}{8953} (\bibinfo{year}{2000}).

\bibitem[{\citenamefont{Tian, Kamenev, and
      Larkin}(2005)}]{kn:tian2005}
Perhaps unexpectedly, with an
appropriately coarse grained density function this principle 
also appears in the `Moyal quantization'
\cite{kn:zachos2002} --- much resembling periodic orbit theory
through the appearance of the diffraction factor similar to the 
factor $\cos(us/\hbar)$ in Eq.~(\ref{eq:dK1med}). 
(Likewise, that the cosine appearing in 
Eq.~(\ref{eq:dK2}) may be also obtained within the
Moyal quantization.) The same principle is described for the
kicked rotator in
\bibinfo{author}{\bibfnamefont{C.} \bibnamefont{Tian}},
\bibinfo{author}{\bibfnamefont{A.} \bibnamefont{Kamenev}}
\bibnamefont{and}
  \bibinfo{author}{\bibfnamefont{A.} \bibnamefont{Larkin}},
  \bibinfo{journal}{Phys. Rev. B} \textbf{\bibinfo{volume}{72}},
  \bibinfo{pages}{045108} (\bibinfo{year}{2005}).

\bibitem[{\citenamefont{Zachos}(2002)}]{kn:zachos2002}
For a review, see, {\em e.g.}, 
\bibinfo{author}{\bibfnamefont{C.}
 \bibnamefont{Zachos}}, 
\bibinfo{journal}{Int. J. Mod. Phys. A} \textbf{\bibinfo{volume}{17}},
 \bibinfo{pages}{297} (\bibinfo{year}{2002}) and references therein.

\end{thebibliography}
\end{document}